\documentclass{jpsj3}

\title{Change of the Ground State upon Hole Doping Unveiled by Ni Impurity in High-$T_{\rm c}$ Cuprates}

\author{Yoichi Tanabe\thanks{E-mail address: youichi@sspns.phys.tohoku.ac.jp}, Kensuke Suzuki, Tadashi Adachi, Yoji Koike, Takayuki Kawamata$^1$, Risdiana$^{1,2}$, Takao Suzuki$^1$, and Isao Watanabe$^1$}

\inst{Department of Applied Physics, Graduate School of Engineering, Tohoku University, \\6-6-05 Aoba, Aramaki, Aoba-ku, Sendai 980-8579 \\
$^1$Advanced Meson Science Laboratory, Nishina Center for Accelerator-Based Science, RIKEN, 2-1 Hirosawa, Wako 351-0198, Japan \\
$^2$Department of Physics, Padjadjaran University, Jl. Raya Bandung-Sumedang km.21 Jatinangor@Sumedang 45363, Indonesia}

\abst{The electronic ground state in high-$T_{\rm c}$ cuprates where the superconducting state is suppressed by Ni substitution has been investigated in La$_{2-x}$Sr$_x$Cu$_{1-y}$Ni$_y$O$_4$ from the specific heat and muon spin relaxation measurements. It has been found that the ground state changes from a magnetically ordered state with the strong hole-trapping by Ni to a metallic state with the Kondo effect of Ni with increasing hole-concentration. Moreover, the analysis of the results has revealed that a phase separation into the magnetically ordered phase and the metallic phase occurs around the boundary of two phases.}

\kword{hole trapping, Kondo effect, La$_{2-x}$Sr$_x$Cu$_{1-y}$Ni$_y$O$_4$, specific heat, muon spin relaxation, quantum phase transition, phase separation}

\begin{document}
\maketitle

In high-$T_{\rm c}$ cuprates, whether or not a quantum critical point (QCP) resides inside the superconducting (SC) phase on the phase diagram has been an important central issue.
It has been pointed out from nuclear magnetic resonance (NMR) \cite{Alloul, Wuyts, Tallon1}, specific heat \cite{Loram1, Loram2}, electrical resistivity \cite{Wuyts, Williams, Konstantinovic}, Hall coefficient \cite{Hwang}, infrared conductivity \cite{Bernhard}, Raman spectroscopy, \cite{Kendziora} and angle-resolved photoemission spectroscopy \cite{Ding} measurements that the magnitude of the pseudo gap decreases with increasing hole-concentration, $p$, such that the pseudo gap appears to vanish at $p$ $\sim$ 0.19 \cite{Tallon2}.
Moreover, it has been reported that the electrical resistivity in La$_{2-x}$Sr$_x$CuO$_4$ (LSCO), Bi$_2$Sr$_{2-x}$La$_x$CuO$_{6+\delta}$, and YBa$_2$Cu$_3$O$_{7-\delta}$ shows a $T$-linear behavior in a very limited region above $T_{\rm c}$ near the optimally doped regime, indicating the appearance of a quantum-critical non-Fermi-liquid state \cite{Gurvitch, Ando}.
These results suggest that a QCP resides inside the SC phase on the phase diagram in high-$T_{\rm c}$ cuprates.
To confirm this, a study of the ground state inside the SC phase on the phase diagram is required.
In fact, in the case when the SC state is suppressed by the partial substitution of Zn for Cu, it has been found from muon-spin-relaxation ($\mu$SR) measurements in La$_{2-x}$Sr$_x$Cu$_{0.97}$Zn$_{0.03}$O$_4$ that the slowing down of the Cu-spin fluctuations weakens gradually with increasing $p$ in the overdoped regime, suggesting that a phase separation into short-range magnetically ordered and metallic regions occurs \cite{Risdi}.
In the case when the SC state is suppressed by the application of high magnetic field, it has been found that the electrical resistivity is expressed as the sum of $T$-linear and $T^2$ terms at low temperatures in the overdoped regime of LSCO \cite{Cooper}, suggesting that the quantum-critical non-Fermi-liquid state is extended over a wide range of $p$ values or that a phase separation into non-Fermi-liquid and metallic regions occurs in the overdoped regime.

As for Ni-impurity effects, it has been pointed out from neutron scattering \cite{Matsuda, Hiraka}, magnetic susceptibility \cite{Machi}, X-ray absorption fine structure (XAFS) \cite{Hiraka2}, and $\mu$SR measurements \cite{Tanabe} in La$_{2-x}$Sr$_x$Cu$_{1-y}$Ni$_y$O$_4$ (LSCNO) that Ni$^{\rm 2+}$ with the spin quantum number $S$ = 1 tends to trap a hole.
In this case, the effective $S$ value of Ni$^{\rm 2+}$ accompanied by a hole changes from 1 to 1/2, such that Ni substituted for Cu is expected not to disturb the $S$ = 1/2 network of Cu$^{\rm 2+}$ spins markedly.
Moreover, the Ni substitution can easily suppress the superconductivity.
Therefore, the Ni substitution is a good candidate for studying the ground state inside the SC phase on the phase diagram in high-$T_{\rm c}$ cuprates.

In this study, we have carried out systematic measurements of the specific heat and $\mu$SR on partially Ni-substituted LSCNO to elucidate the ground state inside the SC phase on the phase diagram.
The electronic specific heat is a good probe for studying the electronic density of states at the Fermi level in the ground state.
The $\mu$SR is a very sensitive probe for detecting the emergence of a static or quasi-static internal magnetic field due to the development of a magnetically ordered state.
We have observed a gradual crossover of the ground state between a magnetically ordered state in the underdoped regime and a metallic Kondo state in the overdoped regime.
The analysis of the results has revealed that the gradual crossover of the ground state is due to a phase separation into short-range magnetically ordered and metallic regions in the overdoped regime.

%Experimental
Polycrystalline samples of LSCNO with $x$ = 0.08 - 0.30 and $y$ = 0 - 0.10 were prepared by an ordinary solid-state reaction method \cite{Adachi0}.
All the samples were determined to be of the single phase by powder X-ray diffraction measurements.
Electrical-resistivity measurements were also carried out using the four-probe method to determine the quality of the samples and to study the electronic state.
Specific heat measurements were carried out using the thermal-relaxation method at low temperatures down to 0.4 K.
Zero-field (ZF) $\mu$SR measurements were performed at temperatures down to 0.3 K at the Paul Scherrer Institute (PSI) in Switzerland and at the RIKEN-RAL Muon Facility at the Rutherford-Appleton Laboratory in the UK.

%Results

Typical results of the specific heat $C$ in  LSCNO are shown in Fig. 1.
The specific heat of non-superconducting LSCNO at low temperatures is given by
\begin{eqnarray}
\label{eq:HC}
C&=&\gamma T + \beta T^3.
\end{eqnarray} 
The first term represents the electronic specific heat $C_{\rm el}$.
$\gamma$ is the Sommerfeld constant, which is proportional to the electronic density of states at the Fermi level.
The second term represents the phonon specific heat $C_{\rm ph}$.
The variation in $C$/$T$ with $T^2$ follows a straight line in the temperature range of 30 K$^2$ $\leq$ $T^2$ $\leq$ 50 K$^2$ for all the samples, indicating that $C_{\rm ph}$ is proportional to $T^3$ and $C_{\rm el}$/$T$ is constant in this temperature range.
For $x$ = 0.13 and $y$ = 0.05, and $x$ = 0.18 and $y$ = 0.10, $C$/$T$ deviates downward from the straight line and decreases with decreasing temperature at $T^2$ $\leq$ 20 K$^2$.
For $x$ = 0.18 and $y$ = 0.05, and $x$ = 0.23 and $y$ = 0.10, on the other hand, $C$/$T$ deviates upward at $T^2$ $\leq$ 30 K$^2$ and shows a local maximum at a low temperature, indicating that $C_{\rm el}$/$T$ is not constant but changes at low temperatures.

After removing the phonon contribution from the total specific heat, the temperature dependence of ($C$ - $C_{\rm ph}$)/$T$ for LSCNO with $y$ = 0.05 and 0.10 is shown in Figs. 2(a) and 2(b), respectively.
The ($C$ - $C_{\rm ph}$)/$T$ is constant at approximately 5 - 7 K for all the samples.
For $x$ = 0.08 and $y$ = 0.05, and $x$ = 0.13 - 0.18 and $y$ = 0.10, ($C$ - $C_{\rm ph}$)/$T$ decreases with decreasing temperature, indicating a decrease in $C_{\rm el}$/$T$.
For $x$ = 0.13 - 0.18 and $y$ = 0.05, and $x$ = 0.20 and $y$ = 0.10, ($C$ - $C_{\rm ph}$)/$T$ once increases and then decreases with decreasing temperature.
For $x$ = 0.20 and $y$ = 0.05, and $x$ = 0.23 and $y$ = 0.10, ($C$ - $C_{\rm ph}$)/$T$ once increases and then tends to be saturated at low temperatures.
For $x$ = 0.22 - 0.30 and $y$ = 0.05, and $x$ = 0.25, 0.30 and $y$ = 0.10, ($C$ - $C_{\rm ph}$)/$T$ continues to increase with decreasing temperature down to 0.4 K, indicating an increase in $C_{\rm el}$/$T$.
Thus, the change in $C_{\rm el}$/$T$ with decreasing temperature is considered to indicate a gradual crossover from the decrease to the increase with increasing $x$.
For $x$ = 0.08 and $y$ = 0.05, and $x$ = 0.13, 0.15 and $y$ = 0.10, ($C$ - $C_{\rm ph}$)/$T$ shows an upturn at very low temperatures below 0.8 K, which may be the so-called Schottky anomaly often observed in high-$T_{\rm c}$ cuprates \cite{Nohara}.

Here, we define constant values of ($C$ - $C_{\rm ph}$)/$T$ at approximately 5 - 7 K as $\gamma_{\rm HT}$.
Figure 3(a) shows the dependence of $\gamma_{\rm HT}$ on the Sr-concentration $x$ for LSCNO with $y$ = 0.05 and 0.10 together with normal-state values of $\gamma$ in Ni-free LSCO obtained by Momono and Ido.\cite{Momono}
It is found that $\gamma_{\rm HT}$ is almost identical to $\gamma$ at each $x$.
According to XAFS measurements in the underdoped regime of LSCNO by Hiraka $et$ $al$. \cite{Hiraka2}, a hole is bound to the substituted Ni$^{\rm 2+}$ even at 290 K, but it does not appear that this affects the value of $\gamma_{\rm HT}$.
Probably, the hole will be weakly bound to Ni$^{\rm 2+}$ at temperatures above 5 K, leading to a constant $\gamma_{\rm HT}$.

Supposed that a Ni$^{\rm 2+}$ ion traps a hole strongly, the effective hole-concentration $p_{\rm eff}$ is defined as $p_{\rm eff}$ = $x$ - $y$.
Here, we define values of ($C$ - $C_{\rm ph}$)/$T$ for $x$ = 0.13 - 0.30 and $y$ = 0.05, and $x$ = 0.18 - 0.30 and $y$ = 0.10 at 0.4 K as $\gamma_{\rm LT}$.
For $x$ = 0.08 and $y$ = 0.05, and $x$ = 0.13, 0.15 and $y$ = 0.10, ($C$ - $C_{\rm ph}$)/$T$ shows an upturn, which may be due to the Schottky anomaly below 0.8 K.
Thus, we define minimum values of ($C$ - $C_{\rm ph}$)/$T$ for these samples at low temperatures below 5 K as $\gamma_{\rm LT}$.
Figure 3(b) shows the dependence of $\gamma_{\rm LT}$ on $p_{\rm eff}$ for LSCNO with $y$ = 0.05 and 0.10 together with $\gamma$ in Ni-free LSCO \cite{Momono}.
In fact, for $p_{\rm eff}$ $\leq$ 0.10, $\gamma_{\rm LT}$ is almost identical to $\gamma$ at each $p_{\rm eff}$ except for $x$ = 0.20 and $y$ = 0.10, indicating the occurrence of the strong binding of a hole around Ni$^{\rm 2+}$ or the fromation of a Ni$^{\rm 3+}$ state at low temperatures below 5 K.
It has been pointed out that the strong binding of a hole around Ni$^{\rm 2+}$ is realized more easily than the formation of a Ni$^{\rm 3+}$ state, because this system is regarded as a Mott insulator of the charge-transfer type \cite{Hiraka2}.
Accordingly, it is inferred that a hole is strongly bound around Ni$^{\rm 2+}$ at low temperatures, leading to the reduction in the electronic density of states at the Fermi level.
That is, it appears that the state of holes around Ni$^{\rm 2+}$ changes from the weakly bound state observed by XAFS by Hiraka $et$ $al$. \cite{Hiraka2} to the strongly bound state with decreasing temperature in these samples.

For $p_{\rm eff}$ $\geq$ 0.13, on the other hand, $\gamma_{\rm LT}$ is much larger than $\gamma$ at each $p_{\rm eff}$. 
That is, ($C$ - $C_{\rm ph}$)/$T$ increases at low temperatures, as shown in Figs. 2(a) and 2(b).
This behavior is reminiscent of the Kondo effect \cite{Kondo}.
The temperature dependence of ($C$ - $C_{\rm ph}$)/$T$ for $x$ = 0.30 and $y$ = 0.10 at a magnetic field of 9 T is shown in Fig. 2(b). 
At 9 T, the increase in ($C$ - $C_{\rm ph}$)/$T$ is suppressed and a broad peak is observed at approximately 4 K.
In a dilute Kondo system, the increase in $C$/$T$ is typically suppressed by the application of magnetic field so that the temperature dependence of $C$/$T$ shows a peak due to the Zeeman splitting of the magnetic impurity level.
Therefore, the observed field dependence of ($C$ - $C_{\rm ph}$)/$T$ is explained as being due to the Kondo effect.
Moreover, it has been observed that the electrical resistivity for $x$ = 0.23 - 0.30 and $y$ = 0.10 shows a minimum at a low temperature and then logarithmically increases with decreasing temperature, as typically observed in a dilute Kondo system.
Accordingly, in a metallic state with a large value of $p_{\rm eff}$, it is inferred that $S$ = 1 spins of Ni$^{\rm 2+}$ in the $S$ = 1/2 network of Cu$^{\rm 2+}$ spins tend to be screened by holes, leading to a Kondo state, namely, a weakly localized state of holes around Ni$^{\rm 2+}$ in the ground state.

For $x$ = 0.13, 0.15, 0.18 and $y$ = 0.05, and $x$ = 0.20 and $y$ = 0.10, ($C$ - $C_{\rm ph}$)/$T$ once increases below 5 K and then decreases with decreasing temperature.
Moreover, for $x$ = 0.20 and $y$ = 0.05, and $x$ = 0.23 and $y$ = 0.10, ($C$ - $C_{\rm ph}$)/$T$ once increases with decreasing temperature and then tends to be saturated at low temperatures.
These behaviors indicate that the state of holes around Ni$^{\rm 2+}$ changes from the Kondo state, namely, the weakly localized state, to the strongly bound state with decreasing temperature in these samples.

%Discussion
In the underdoped regime of high-$T_{\rm c}$ cuprates, the development of a magnetically ordered state at low temperatures may be correlated with the localization of holes \cite{Kitazawa}.
Therefore, we have performed $\mu$SR measurements to investigate the magnetic transition temperature $T_{\rm N}$ in LSCNO.
The $p_{\rm eff}$ dependence of $T_{\rm N}$ is shown in Fig. 4.
$T_{\rm N}$ is described in our previous paper \cite{Adachi}.
Both the onset temperature of the strongly bound state of a hole around Ni$^{\rm 2+}$, $T_{\rm SB}$, and the onset temperature of the Kondo effect, $T_{\rm Kondo}$, are also plotted in Fig. 4.
The $T_{\rm SB}$ is defined as the onset temperature of the decrease in ($C$ - $C_{\rm ph}$)/$T$ with decreasing temperature.
For $x$ = 0.20 and $y$ = 0.05, and $x$ = 0.23 and $y$ = 0.10, $T_{\rm SB}$ is defined as the onset temperature of the saturation of ($C$ - $C_{\rm ph}$)/$T$ with decreasing temperature.
The $T_{\rm Kondo}$ is defined as the onset temperature of the increase in ($C$ - $C_{\rm ph}$)/$T$.
It is found that $T_{\rm N}$ decreases with increasing $p_{\rm eff}$ and appears to vanish at $p_{\rm eff}$ = 0.15 - 0.20.
Values of $T_{\rm N}$ for 5 and 10$\%$ Ni-substituted LSCNO samples are in rough agreement with those for Ni-free LSCO, suggesting that the really effective hole-concentration in each sample is well described by $p_{\rm eff}$.
That is, the magnetically ordered state in each $p_{\rm eff}$ of LSCNO corresponds to the magnetically ordered state in each $x$ of LSCO.
Moreover, it is found that $T_{\rm N}$ roughly coincides with $T_{\rm SB}$ in each $p_{\rm eff}$, indicating that the strong hole-trapping by Ni$^{\rm 2+}$ at low temperatures is strongly correlated with the stabilization of the magnetically ordered state.
The coincidence between $T_{\rm N}$ and $T_{\rm SB}$ may not be accidental, because the effective $S$ value of Ni$^{\rm 2+}$ trapping a hole strongly changes from 1 to 1/2 such that the $S$ = 1/2 network of Cu$^{\rm 2+}$ spins is restored to some extent.
Moreover, the strong binding of a hole by Ni$^{\rm 2+}$ is expected to reduce the frustration effect between Cu$^{\rm 2+}$ spins.
That is, it appears that the gain of the magnetic free energy due to the emergence of the magnetically ordered state at each $p_{\rm eff}$ induces the strong hole-trapping by Ni$^{\rm 2+}$.
In a regime with large $p_{\rm eff}$ values, on the other hand, a Kondo state emerges at low temperatures, as mentioned above.
The Kondo state is regarded as a sign of a metallic ground state.
Therefore, the present result of the change of the electronic state around Ni$^{\rm 2+}$ from the strongly bound state of a hole around Ni$^{\rm 2+}$ to the Kondo state due to Ni$^{\rm 2+}$ spins in LSCNO indicates the change of the ground state from a magnetically ordered state to a metallic state inside the SC phase on the phase diagram in LSCO with increasing $p$.

Finally, we discuss possible scenarios for understanding the change of the ground state in LSCNO.
For $x$ = 0.22, 0.23 and $y$ = 0.05, ($p_{\rm eff}$ = 0.17, 0.18), ($C$ - $C_{\rm ph}$)/$T$ logarithmically increases toward 0 K, as shown in Fig. 2(a).
These results are consistent with the quantum critical behavior of a two-dimensional antiferromagnet \cite{Moriya}.
Moreover, $T_{\rm N}$ appears to vanish at $p_{\rm eff}$ = 0.15 - 0.20.
Therefore, one might be tempted to take the change of the ground state from the magnetically ordered state to the metallic state as the occurrence of a quantum phase transition at $p_{\rm eff}$ = 0.17 - 0.20.
However, the specific heat results for $x$ = 0.18, 0.20 and $y$ = 0.05, and $x$ = 0.20, 0.23 and $y$ = 0.10 indicate that the electronic state exhibits a crossover from a Kondo state to a magnetically ordered state with decreasing temperature.
Therefore, these results are not consistent with the simple scenario of a quantum phase transition at $p_{\rm eff}$ = 0.17 - 0.20 and instead bear out a scenario based on the occurrence of an inhomogeneous electronic state.
In fact, the magnetic susceptibility measurements have revealed that a microscopic phase separation into SC and normal-state regions occurs in the overdoped regime of LSCO \cite{Tanabe1, Tanabe2}.
Supposed that a phase separation into short-range magnetically ordered and metallic regions occurs at around the boundary of the two phases, the temperature dependence of ($C$-$C_{\rm ph}$)/$T$ is represented as the sum of the decrease in ($C$-$C_{\rm ph}$)/$T$ due to the strong hole trapping by Ni$^{\rm 2+}$ and the increase in ($C$-$C_{\rm ph}$)/$T$ due to the Kondo effect.
In this case, $\gamma_{\rm LT}$ is expected to be larger than the $\gamma$ of Ni-free LSCO at each $p_{\rm eff}$.
In fact, values of $\gamma_{\rm LT}$ for $x$ = 0.18, 0.20 and $y$ = 0.05, and $x$ = 0.20, 0.23 and $y$ = 0.10 shown in Fig. 3(b) are larger than those of the $\gamma$ of Ni-free LSCO at each $p_{\rm eff}$.
Therefore, it is likely that the quantum phase transition does not occur at a single point on the phase diagram but is modified to be crossover-like owing to the phase separation into short-range magnetically ordered and metallic regions in the overdoped regime of LSCNO.
This scenario is consistent with the results inside the SC phase on the phase diagram of LSCO obtained through the Zn substitution \cite{Risdi} and the application of high magnetic field \cite{Cooper}.

%Summary
In conclusion, we have succeeded in unveiling a gradual crossover of the ground state between a magnetically ordered state and a metallic state inside the SC phase on the phase diagram using the Ni impurities in LSCO.
The present results suggest that QCP is modified to be crossover-like owing to the phase separation into short-range magnetically ordered and metallic regions in LSCO.
This scenario based on the phase separation in the overdoped regime is consistent with the scenario unveiled by the Zn substitution and the application of high magnetic field in LSCO.

%Acknowledge
We are grateful to Y. Shimizu, H. Tsuchiura, M. Ogata, T. Tohyama, and K. Yamada for helpful discussions.
We also thank A. Amato and R. Sheuermann at PSI for their technical support in the $\mu$SR measurements. 
The $\mu$SR measurements at PSI were partially supported by the KEK-MSL Inter-University Program for Oversea Muon Facilities and also by the Global COE Program gMaterials Integration (International Center of Education and Research), Tohoku University,h MEXT, Japan.
One of the authors (Y. T.) was supported by the Japan Society for the Promotion of Science.

\begin{figure}[htbp]
\begin{center}
\includegraphics[width=1.0\linewidth]{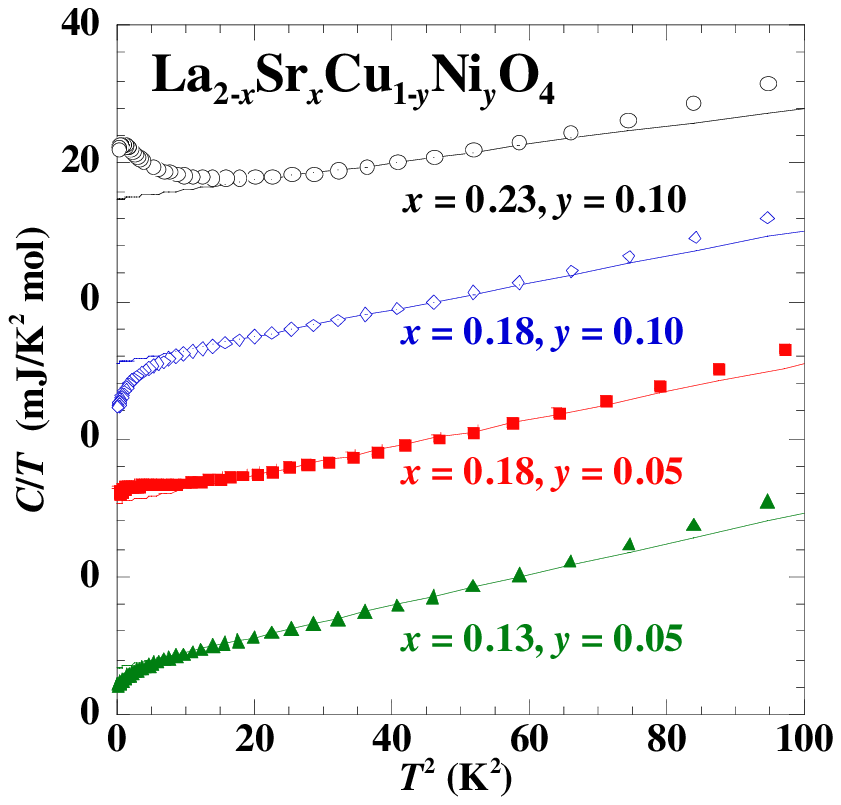}
\end{center}
\caption{(color online) Temperature dependence of the specific heat $C$ for typical values of $x$ in La$_{2-x}$Sr$_x$Cu$_{1-y}$Ni$_y$O$_4$ with $y$ = 0.05 and 0.10 plotted as $C$/$T$ vs $T^2$. Solid lines indicate the best-fit results obtained using eq. (\ref{eq:HC}) in the temperature range of 30 K$^2$ $\leq$ $T^2$ $\leq$ 50 K$^2$.}
\end{figure}

\begin{figure}[htbp]
\begin{center}
\includegraphics[width=1.0\linewidth]{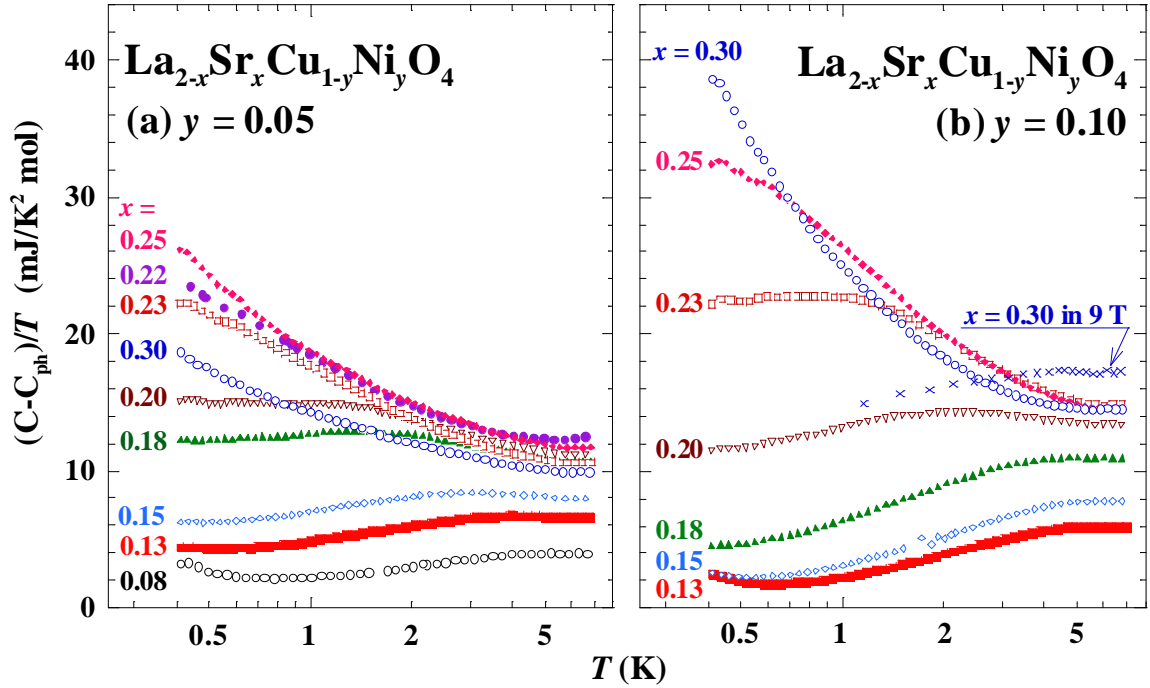}
\end{center}
\caption{(color online) Temperature dependence of ($C$ - $C_{\rm ph}$)/$T$ for La$_{2-x}$Sr$_x$Cu$_{1-y}$Ni$_y$O$_4$ with (a) $y$ = 0.05 and (b) $y$ = 0.10 at zero field. The $C_{\rm ph}$/$T$ is obtained from the best-fit results obtained using eq. (\ref{eq:HC}) in the temperature range of 30 K$^2$ $\leq$ $T^2$ $\leq$ 50 K$^2$ and then is subtracted from $C$/$T$ to estimate the electronic contribution to the specific heat. Data of $x$ = 0.30, $y$ = 0.10 at a magnetic field of 9 T are also indicated by crosses.}
\end{figure}

\begin{figure}[htbp]
\begin{center}
\includegraphics[width=1.0\linewidth]{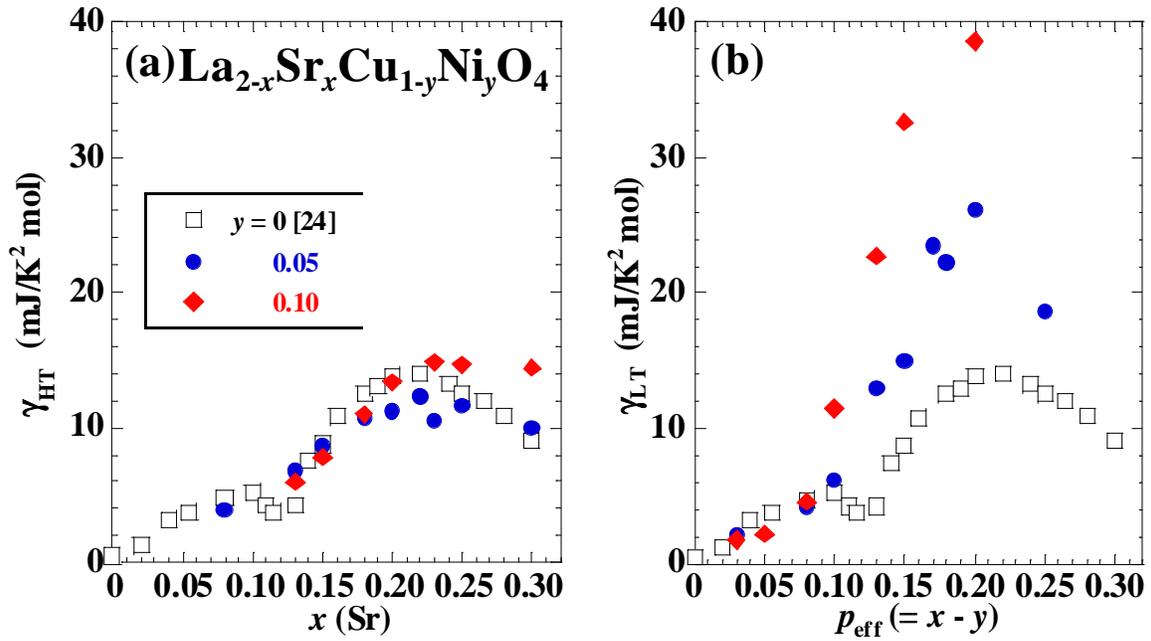}
\end{center}
\caption{(color online) (a) Sr-concentration $x$ dependence of $\gamma_{\rm HT}$ for La$_{2-x}$Sr$_x$Cu$_{1-y}$Ni$_y$O$_4$ with $y$ = 0.05 (closed circles) and 0.10 (closed diamonds). (b) Effective hole-concentration $p_{\rm eff}$ dependence of $\gamma_{\rm LT}$ for La$_{2-x}$Sr$_x$Cu$_{1-y}$Ni$_y$O$_4$ with $y$ = 0.05 (closed circles) and 0.10 (closed diamonds). $\gamma_{\rm HT}$ and $\gamma_{\rm LT}$ are described in the text. Normal-state values of $\gamma$ in Ni-free La$_{2-x}$Sr$_x$Cu$_{1-y}$Ni$_y$O$_4$ with $y$ = 0 (open squares) obtained by Momono and Ido \cite{Momono} are also plotted in (a) and (b).}
\end{figure}

\begin{figure}[htbp]
\begin{center}
\includegraphics[width=1.0\linewidth]{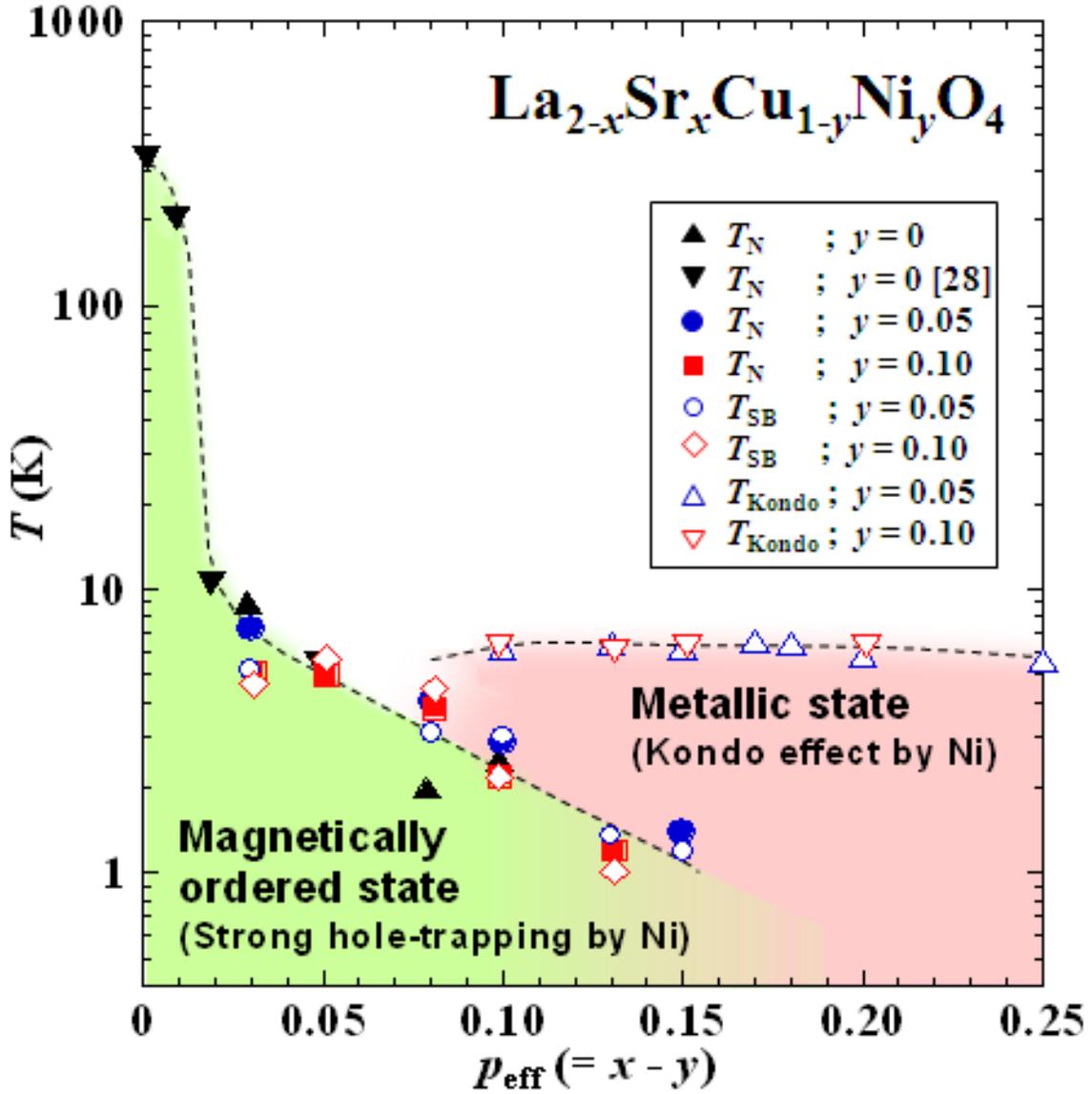}
\end{center}
\caption{(color online) Electronic phase diagram for La$_{2-x}$Sr$_x$Cu$_{1-y}$Ni$_y$O$_4$ obtained from specific heat and $\mu$SR measurements. Effective hole-concentration $p_{\rm eff}$ dependence of the magnetic transition temperature $T_{\rm N}$ estimated from the $\mu$SR measurements for La$_{2-x}$Sr$_x$Cu$_{1-y}$Ni$_y$O$_4$ with $y$ = 0 (closed triangles, closed inverted triangles \cite{Budnick}), 0.05 (closed circles), and 0.10 (closed squares). The onset temperature of the strong localization of a hole around Ni$^{\rm 2+}$, $T_{\rm SB}$, for $y$ = 0.05 (open circles) and 0.10 (open diamonds), and the onset temperature of the Kondo effect, $T_{\rm Kondo}$, for $y$ = 0.05 (open triangles) and 0.10 (open inverted triangles) estimated from the specific heat measurements are also plotted. Dashed lines serve as visual guides.}
\end{figure}

\end{document}